%Paper: astro-ph/9301012
%From: p686bin@mpifr-bonn.mpg.de (Biman Nath)
%Date: Fri, 29 Jan 93 16:06:35 +0100

\newcount\fcount \fcount=0
\def\ref#1{\global\advance\fcount by 1 \global\xdef#1{\relax\the\fcount}}

\def\la{\lower.5ex\hbox{$\; \buildrel < \over \sim \;$}}
\def\ga{\lower.5ex\hbox{$\; \buildrel > \over \sim \;$}}

\magnification=\magstep1

\raggedbottom
\footline={\hss\tenrm\folio\hss}

\tolerance = 30000

\baselineskip=5.mm plus 0.1mm minus 0.1mm
\centerline {\bf A POSSIBLE FOREST OF EMISSION LINES FROM PROTO-GALAXIES}
\vskip 0.5cm
\centerline {\bf Biman B. Nath$^1$ \& David Eichler$^2$}
\centerline{$^1$Max Planck Institut f\"ur Radioastronomie,
Auf dem H\"ugel 69, 1 D-5300 Bonn}
\centerline{$^2$Department of Physics, Ben-Gurion University,
Beer-Sheva, 84105 Israel}

\vskip 2cm
\centerline {\bf Summary}

The possibility of detecting proto-galaxies in the UV band is
pointed out, assuming galaxy formation occured at z $\sim 5-6$.
It is shown that the diffuse gas in collapsing galaxy sized objects with
temperatures $\sim 10^{6\pm0.5}$ K, and with a modest amount
of metallicity, should copiously produce emission lines from highly
ionized Iron atoms. The expected luminosity from models of
galaxy formation is compared with
the sensitivity of HST.


%\vfill\eject
\bigskip
\noindent
{\bf 1. Introduction }
\medskip
The search for galaxies at birth is one of the central goals
of modern observational cosmology. The work of Partridge and Peebles (1967)
was among the first theoretical studies on how galaxies might appear
at their forming stage. They calculated the
luminosity of galaxies in hydrogen Ly$\alpha_+$ when the bulk of their stars
were forming at $z\sim 10-30$ and predicted their appearances as diffuse
red objects. Meier (1976) found the hydrogen Ly$\alpha$ luminosity to be of the
order of $\sim 10^{41}$ ergs/s and that these objects would
appear blue at much lower redshifts. Other
studies on the possible appearances of proto-galaxies, from various
mechanisms for emission, have in general resulted in such order of
magnitude of brightness or bigger. Shull and Silk (1979) considered
emission from
supernova remnants inside proto-galaxies in UV wavelengths. Baron and White
(1987) considered models of dissipative collapse resulting in slow
star formation and, thus, with dimmer and more extended appearances. Various
authors have also calculated the radiation from the dust from primeval
star formation in infrared wavelengths. A recent paper that reviews various
aspects
of the observability of proto-galaxies is that of Djorgovsky and
Thompson (1992)(hereafter referred to as DT92).

Observers usually consider the phase of the bulk of star formation
as defining the galaxy formation phase. However, the continuum
radiation from star formation is expected to be relatively flat
(except for the Lyman break at $912\AA$), and, therefore,
lacking any information about the redshift. One then looks for emission
line signatures, such as, hydrogen Ly$\alpha$. A variety of Ly$\alpha$ objects
have already been discovered at moderately high redshifts($z\sim 1.8-3.8$), but
their
identification as genuine forming galaxies is suspect because most of them
seem to be associated with AGN sources (DT92).

One uncertainty in searches for proto-galaxies lies in selecting
the wavelength for detection and stems from
our ignorance of the epoch of galaxy formation. Though the above mentioned
objects (at $z\sim 2-3$) intrigue us with the possibility of being real
proto-galaxies, studies suggest that at least some fraction of these systems
formed at $z>5$ (Hamilton (1985), Gunn {et. al} (1986)).
Moreover, the existence of a large number of radio galaxies at $z>3$ and
the absence of a sharp quasar number density cut off out to $z \sim5$,
argue for a higher galaxy formation redshift than has been searched
so far (i.e., $z\la 5$).

Here we consider the possibility of a definite spectral signature
from highly ionized Fe atoms in the diffuse gas inside forming galaxies.
It is very likely that the diffuse gas in the proto-galaxies,
possibly coexisting with the first stars and hence with some metalicity,
reached temperatures of the order of $10^{6\pm0.5}$K, either from
virialization due to the gravitational potential of the halo or because of some
other mechanisms. In that case, as we show later, various emission lines of
Fe atoms stand the chance of being the most luminous ones and of giving
specific information about the redshift.

We begin by stating our assumptions and indicating the motivations for
considering temperatures in the range $\sim 10^{6.0\pm0.5}$ K for the gas.

\bigskip
\noindent
{\bf 2. Luminosity in Fe lines}
\medskip
{\it (a) Temperature of the diffuse gas}: For simplicity we assume
an $\Omega=1$ and $h=1$ universe.
Consider a cloud of gas inside the halo of dark
matter. If we assume that the halo of dark matter can be modeled as a
spherically
symmetric isothermal sphere (of temperature $T$), then hydrostatic
equilibrium yields, $T=1.44\times 10^6
({V_c\over 200 km/s})^2 $ K (with $\mu=0.59$).
Here,
$V_c=(GM(r)/r)^{0.5}$ is the circular velocity.

Thus, if the gas is at the virial
temperature of the halo potential, then with the circular
velocities of the present day spirals like Milky Way, the
temperature would be of the order of $\sim 10^6$K.
Merger of haloes and the subsequent collisions between
subgalactic fragments could also have released a large amount of gas
at high temperature.
Collisions between clouds with velocities $\sim 200$ km/s
during the merger event could result in temperatures $\sim 10^6$K.
Moreover, as Shull and Silk (1979) considered, supernova
remnants from the first bursts of star formation could lead
to high temperatures for the diffuse gas.

{\it (b) Luminosity}:
First we calculate the cooling function
of a gas with temperature $\sim 10^6$ K due to emission of various
atomic lines using the results of Gaetz and Salpeter (1983). The most
important lines at this temperature are [Fe VII] $\lambda\lambda 166.2
,\>176.9\AA$, [Fe VIII]
$\lambda\lambda 167.9,\>168.7,\>185.8 \AA$, [Fe IX] $\lambda 171.1 \AA$,
[Fe X] $\lambda\lambda 174.5,\> 178.3,\> 186.3\AA$, [Fe XI] $\lambda
\lambda 180.7,\> 189.3 \AA$, [Fe XII] $\lambda\lambda 189.7,\>190.1,
\>194.1,\>196.1$.
Collisional ionization equilibrium is expected to hold, and the
deviation in the ionization fraction of highly ionized Fe atoms
due to non-equilibrium processes
negligible for the above range of
temperatures (Schmutzler and Tscharnuter 1993).
In the possibility of runaway cooling at $T> 10^6$ K,
if the heating sources continue to keep the gas hot,
ionization equilibrium may collapse, but the gas
soon reaches a temperature $\sim 10^6$ K, where the cooling
is maximum.
The emission in line photons is more or comparable
to that due to thermal bremsstrahlung at these wavelengths for metallicities
$Z \ga 0.075$ and we only consider the former.
The constructed cooling function is shown in fig. 1.
Fig. 2 shows the relative intensities of various
lines for different temperatures.

If the density profile of the baryonic gas (of radius $R$) is taken to be
$n\propto r^{-\alpha}$, with uniform metallicity,
one then easily calculates the total luminosity of the cloud of gas
in these lines. At $10^6$ K, e.g., one gets, for $\alpha < 1.5$,
$$
L_{Fe} \approx 5 \times 10^{44}
\Bigl({M_{gas}\over10^{10} M_{\odot}}\Bigr)^2\Bigl({Z\over Z_{\odot}}\Bigr)
\Bigl({1\over \mu}\Bigr)^2 {(3-\alpha)^2\over 3-2\alpha}
\Bigl({R\over 5\> {\rm kpc}}\Bigr)^{-3} \; {\rm erg/s}
\eqno(2.1)
$$

As the brightness
does not strongly depend on $\alpha$, we shall use a value of $1$. Considering
the extent of the cloud till $R\sim 5$ kpc, we have calculated
the luminosity of a few cases. Smaller values of $R$ would correspond
to higher contours toward the center, but the very central region ($\ll 1$
kpc) is likely to be in a very different state of affairs than the outer few
kiloparsecs.
Thresholds of detection with the FOC camera aboard HST
are shown in the $M_{gas}-T$ space in fig. 3, for clouds with $R=5$ kpc.
The filter F140W with $\lambda=1360\AA$ and $\Delta\lambda=298\AA$
seems to be suitable for the above lines from a redshift of $z\sim5-6$.
For larger values of $R$, the integration time to detect the same mass of
gas inside $R$ would scale as $ \sim R^{-6}$.

Intensities of the individual lines, shown in fig. 2,
indicate that the most prominent lines contribute to the order of $1/2$
of the total luminosity. Once any candidate proto-galaxy has been
detected with imaging, then one could integrate longer on such objects
in the spectroscopic mode to detect the prominent lines in the forest.
As was explained earlier, the spectral signature would bear information
about the redshift of its origin. Imaging and subsequent follow up
with spectroscopic details could, therefore, result in detection and
identification
of such proto-galactic objects.
Fig. 2 shows that at any temperature there are at least two lines
with power (erg cm$^3$ /s) $\ga 10^{-24} n_e n_H$ ($Z=0.1Z_{\odot}$).
The threshold contours for detecting the prominent lines
are shown in fig. 4.
After the optics in HST is
corrected, the sensitivity is expected to go up and fainter galaxies
could be detected then.

The opacity of the lines at the line center can be written as
$$\eqalignno{
\tau_l&=1.2 \times 10^{-18} X_4 T_6^{-0.5} \theta_i f
\lambda_{1000} \mu^{1.5} N&\cr
\quad&=1.8 \times 10^{-19} \Bigl({Z\over Z_{\odot}}\Bigr)T_6^{-0.5} \theta_i f
\lambda_{1000} N
&(2.2)\cr}
$$
where $10^{-4} X_4$ is the fractional abundance of the element (here, Fe),
$\theta_i$ is the fractional abundance of the ion, $f$ is the line
oscillator strength, $\lambda_{1000}$ is the wavelength at line center
in units of $1000 \AA$, and $N$ is the column density. We wrote $X_4
=10^{-.5} ({Z\over Z_{\odot}})$ and $\mu=0.6$ for the second expression above.
Larger clouds with $N\sim 10^{22}$ will be marginally
opaque to the most dominant lines for $({Z\over Z_{\odot}})
\sim0.1$. But the collisional deexcitation rate is not high enough to
render the lines thermalized.
In the two level atom model, the line cooling function is
$q={n_l E c_l \exp(-E/kT)\over 1+\gamma}$,
where $\gamma_l=2 n_e c_l(1+\tau_l)
/(A_l)$ is the escape parameter, $A_l$ is the Einstein's $A$
coefficient, $c_l$ is the collisional deexcitation rate of the line,
and $n_l=10^{-4}X_4 n_e \theta_i$.
Using the collision strengths of the lines from Gaetz and Salpeter (1983),
one readily finds that for the proto-galaxies, $\gamma_l\ll1$ and
most of the radiation should emerge from the cloud.

\bigskip
\noindent
{\bf 3. Comparison with other predictions}
\medskip
We can compare the above luminosity with the predicted luminosities
of proto-galactic objects in the literature in different wavelengths.
Meier (1976) concluded
that a star formation rate of $1.0 $ M$_{\odot}$ yr$^{-1}$ in
a primeval galaxy produces a luminosity, which is approximately constant
longward of the Lyman break and is equal to $2.6 \times 10^{27}$ ergs
s$^{-1}$ Hz$^{-1}$. For hydrogen Ly$\alpha$ line, this leads to a
luminosity of
$
\sim 6.7 \times 10^{42} \Bigl({SFR\over 1\> M_{\odot}/yr.}\Bigr)
$ erg/s.
Baron and White (1987)
calculated the Ly$\alpha$ brightness of proto-galaxies at
redshifts of $z\sim 2$ for CDM model. In their model, the collapse of a
galaxy is a highly inhomogeneous process, leading to cloud collisions and
star formation behind the shock fronts.
Their typical example
of a proto-galaxy had two bursts of
star formation of $\sim 70, 12 (\times 10^9 {\rm yr}/t_{coll})$
M$_{\odot}$ yr$^{-1}$ for a final
galaxy with $10^{11}$ M$_{\odot}$ of stars, where $t_{coll}$ is the collapse
time of a galaxy.

Shull and Silk (1979) calculated the UV luminosity of proto-galaxies
arising from shock heated gas due to supernova remnants following a burst
of star formation. With $N_s$ as the rate of supernovae per year and
$n$ as the mean density of particles, they calculated hydrogen Ly$\alpha$
luminosity as
$
3.2 \times 10^{43} n^{-0.5} N_s
$ erg/s.
Luminosity in the band with $\lambda <228 \AA$,
was calculated to be
$10^{41} n^{-0.5} N_s$ erg/s, with $N_s=$few .
\bigskip

\noindent
{\bf 5. Discussion on observability}
\medskip

Finally we discuss a few aspects of the observability of these proto-galactic
objects in UV.

{\it (a) Number density of objects}: Recently White and
Frenk (1991) have presented expressions for the fraction of matter
which is in haloes of a given circular velocity at a certain
redshift in the CDM model.
One must note that various nonlinear processes and the effects
of mergers are yet to be incorporated in a realistic manner. For
a biasing factor $b=1.5$, haloes with $V_c=200$ km/s peak at
$z=4$ with a comoving density of $8.4 \times 10^{-4}$ {\rm Mpc}$^{-3}$.
At $z=6$, the corresponding density is $4.4 \times
10^{-4}$ {\rm Mpc}$^{-3}$. The virialized region of these haloes have
a mass $1.8 \times 10^{11}$ M$_{\odot}$, and radius $76.1$ kpc
(baryons would be concentrated toward the center). They defined virialized
part of the halo as the region within which the mean
overdensity is 200.

The comoving volume can be written as
$
V=4({c\over H_o})^3 (1+z)^{-1.5} \lbrack 1-(1+z)^{-0.5}\rbrack ^2 \theta^2
\Delta z
$
for $\Omega=1$. For $z=5$, this gives $V\sim 2.175 \times 10^{2}
\theta^2 ({\rm arcmin}) \lbrack{\Delta \lambda\over \lambda}\rbrack
(h^{-1} {\rm Mpc})^3$. Turning this over, a comoving
density of $4.4 \times 10^{-4}$ {\rm Mpc}$^{-3}$ means $\sim 0.1$ haloes per
arcminute square, for $\lbrack{\Delta \lambda\over \lambda}\rbrack\sim 1$
and $h=1$.
It is a small number but not prohibitively low (the field of view
of FOC in F/48 mode
is $\sim 0.5$ arc minute square).

In the HDM model and the explosion scenario, galaxies form earlier than in CDM
model. In the explosion model, the expanding shells would have been detectable
in UV but for their large size. The intensity per unit area would
be too small to be detected.

{\it (b) Effect of Dust}: It has been suggested that the reason for null
detection
of proto-galaxies in recent Lyman $\alpha$ searches is obscuration by dust in
those objects.
However, Djorgovsky
and Thompson (DT92) have used the
COBE limit on the sub-mm background to argue against completely obscured
star formation. Furthermore, York and Meyer (1989),
and Fall et. al (1989) found that distant damped Lyman $\alpha$
systems are both dust and metal poor (also see Pettini
{et. al} (1990)).
van den Bergh (1990) concluded that
the effect of dust should not be important as long as  $\lbrack Fe/H\rbrack
\la -1.0$.

{\it (c) Intervening clouds}: Absorption by intervening clouds with high column
density is probably the most worrisome aspect concerning observability of
the Fe emission lines in  the UV. Lyman$\alpha$ clouds, with a
column density in HI $\ga 10^{18}$ cm$^{-2}$, are numerous
and perhaps cover a significant portion of the sky.
It has been estimated (Sargent 1987) that 50\% of randomly chosen lines of
sight
intercept a cloud which absorbs 99\% or more of the flux at the Lyman limit
at a redshift of $z=2$. The fraction rises to 90\% by a redshift of $z=4$.
The chance, therefore, of discovering proto-galaxies at high redshift
in UV may appear to be slim, but considering the importance and
novel impact of such a discovery, such a search would seem to be worthwhile.

\bigskip
\noindent
{\bf Conclusions}
\medskip

We find that photons from highly ionized Fe atoms in the diffuse gas in
collapsing galaxies, with temperatures $\sim 10^{6\pm 0.5}$ K,
should be detectable at redshifts $z\sim 5-6$ in the
UV by HST with a few hours of integration. We suggest spectroscopic
follow up of the candidate objects from imaging which would elicit
emission lines fixing the redshift of their origin.

\bigskip
\centerline{\bf Acknowledgements}
We thank Drs. T. Schmutzler, V. Trimble and A. Wilson for valuable
discussions, Dr. A. Robinson for comments as the referee,
and Warren Hack for help in running FOCSIM. This work
formed part of BN's PhD thesis at the University of Maryland.
\medskip
\centerline{\bf References}

\parskip=0pt
\parindent=0pt
\medskip

{\obeylines
1. Baron, E., White, S. D. M. 1987. {\it ApJ}, {\bf 322}, 525.
2. Blumenthal, G. {\it et. al} 1984. {\it Nature}, {\bf 311}, 517.
3. Djorgovsky, S., Thompson, D. 1992. {\it I.A.U. Symp. 149}, Reidel, Holland.
4. Fall, S.M., Pei, Y., McMahon, R. 1989. {\it ApJLett.}, {\bf 341}, L5.
5. Hamilton, D. 1985. {\it ApJ}, {\bf 297}, 371.
6. Gaetz, T, Salpeter, E. 1983. {\it ApJ. Suppl.}, {\bf 52}, 155.
7. Gunn, J., Hoessel, J., Oke, J 1986. {\it ApJ}, {\bf 306}, 30.
8. Meier, D. 1976. {\it ApJ}, {\bf 203}, L103.
9. Partridge, B., Peebles, P. J. E. 1967. {\it ApJ}, {\bf 147}, 868.
10. Pettini, M., Boksenberg, A., Hunstead, R. 1990. {\it ApJ}, {\bf 348}, 48.
11. Sargent, W. 1987. {\it I.A.U. Symp. 124}, Reidel, Holland.
12. Schmutzler, T., Tscharnuter, W. 1993. preprint.
13. Shull, J. M., Silk, J. 1979. {\it ApJ}, {\bf 234}, 427.
14. Shull, J. M., Van Steenberg, M. 1982. {\it ApJ. Suppl.}, {\bf 48}, 95.
15. van den Bergh, S. 1990. {\it PASP}, {\bf 102}, 503.
16. White, S. D. M., Frenk, C. S. 1991. {\it ApJ}, {\bf 379}, 52.
17. York, D. G., Meyer, D. 1989. {\it Cosmic Abundances of Matter,
AIP Conference Proceedings no. 183}, ed. C. J. Waddington, AIP, New York.}

\vfill\eject
Figure 1. Cooling function for the Fe lines referred to in the text for
$Z=0.1Z_{\odot}$. $P n_e n_H$ is the power radiated per unit volume of
the gas. Collision strengths for individual lines are taken from
Gaetz and Salpeter (1983) and equilibrium fractional abundances for
different ions from Shull and Van Steenberg (1982).
\medskip

Figure 2. Relative intensities of individual lines at $T=10^{5.7}$K
(dashed line), $10^6$ K (solid line) and $10^{6.3}$ K (dotted line).
Metalicity is that of solar abundance and $P n_e n_H$ is the power
radiated per unit volume.
\medskip

Figure 3: Contours of threshold detection of gas clouds of radius
$5$ kpc, as functions of redshift, metallicity and temperature. The
sensitivity used is that of the FOC aboard HST (integration
time $5$ hours, $S/N=5$). Solid and dashed lines refer to metallicities
$Z=0.1, 0.075
Z_{\odot}$ respectively. Luminosity distance and angular
size was calculated for an $\Omega=1, h=1$ universe.

\bigskip
Figure 4: Contours of threshold detection of individual lines
by F/96 FUVOP with integration time $20$ hrs and $S/N=5$. Radius
of gas clouds is taken equal to $5$ kpc and metallicity
$Z=0.1Z_{\odot}$ in an $\Omega=1, h=1$ universe.

\end